\newcommand{\NTENP}{[Ni($N,N'$-bis(3-aminopropyl)propane-1,3-diamine($\mu$-NO$_2$)]ClO$_4$}

\documentclass[aps,prb,twocolumn,superscriptaddress,showpacs]{revtex4}
\usepackage{graphicx}
\usepackage{natbib}
\usepackage{bm}

\begin{document}

\def\be{\begin{equation}}
\def\ee{\end{equation}}
\def\ba{\begin{eqnarray}}
\def\ea{\end{eqnarray}}

\title{Distribution of exchange energy in a bond-alternating $S=1$ quantum spin chain }

\author{A. Zheludev}
\affiliation{Condensed Matter Sciences Division, Oak Ridge
National Laboratory, Oak Ridge, TN 37831-6393, USA.}
 \email{zheludevai@ornl.gov}
 \homepage{http://neutron.ornl.gov/~zhelud/}

\author{T. Masuda}
\affiliation{Condensed Matter Sciences Division, Oak Ridge
National Laboratory, Oak Ridge, TN 37831-6393, USA.}

\author{B. Sales}
\affiliation{Condensed Matter Sciences Division, Oak Ridge
National Laboratory, Oak Ridge, TN 37831-6393, USA.}

\author{D. Mandrus}
\affiliation{Condensed Matter Sciences Division, Oak Ridge
National Laboratory, Oak Ridge, TN 37831-6393, USA.}

\author{T. Papenbrock}
\affiliation{Department of Physics and Astronomy, University of Tennessee,
Knoxville, TN 37996-1200, USA.}
\affiliation{Physics Division, Oak Ridge
National Laboratory, Oak Ridge, TN 37831-6373, USA.}

\author{T. Barnes}
\affiliation{Department of Physics and Astronomy, University of Tennessee,
Knoxville, TN 37996-1200, USA.}
\affiliation{Physics Division, Oak Ridge
National Laboratory, Oak Ridge, TN 37831-6373, USA.}

\author{S. Park}
\affiliation{NIST Center for Neutron Research, National Institute
of Standards and Technology, Gaithersburg, MD 20899, USA.}

\affiliation{Department of Materials Science and Engineering,
University of Maryland, College Park, MD 20742, USA.}

\date{\today}
\begin{abstract}
The quasi-one-dimensional bond-alternating $S=1$ quantum
antiferromagnet \NTENP\ (NTENP) is studied by single crystal
inelastic neutron scattering. Parameters of the measured
dispersion relation for magnetic excitations are compared to
existing numerical results and used to determine the magnitude of
bond-strength alternation. The measured neutron scattering
intensities are also analyzed using the 1st-moment sum rules for
the magnetic dynamic structure factor, to directly determine the
modulation of ground state exchange energies. These independently
determined modulation parameters characterize the level of spin
dimerization in NTENP. First-principle DMRG calculations are used
to study the relation between these two quantities.
\end{abstract}


\maketitle
\section{Introduction}

\begin{figure}
 \includegraphics[width=3.3in]{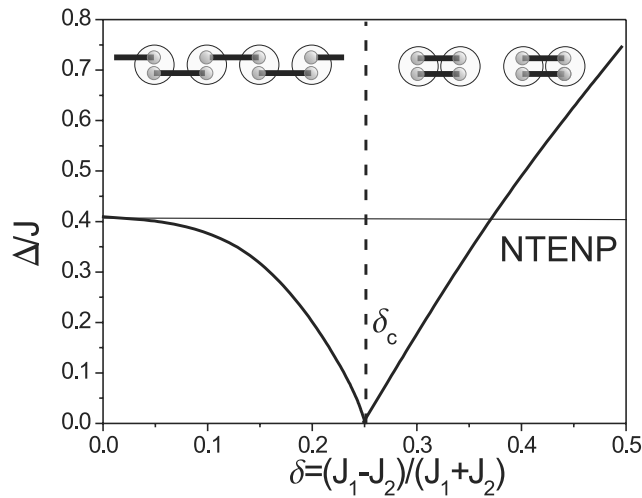}
 \caption{\label{phase} Field dependence of energy gap in a bond-alternating $S=1$ quantum
 antiferromaget, based on numerical results by S. Yamamoto, Ref.~\protect\onlinecite{Yamamoto95}.
 A quantum critical point (dashed line) separates the Haldane phase, similar to the VBS state (left inset)
 from the dimerized state (right inset). The
 possible location of NTENP is indicated by the thin horizontal line.}
\end{figure}

Integral antiferromagnetic (AF) spin chains are best known for
having an exotic spin liquid ground state with a  characteristic
gap in the magnetic excitation
spectrum.\cite{Haldane83,Haldane83-2} The Haldane gap has been
subject to intensive theoretical and experimental studies for the
last two decades, and is by now very well characterized and
understood. The spin wave function of the Haldane ground state is
not known exactly, but is similar to the easy to visualize Valence
Bond Solid (VBS) state.\cite{Affleck89} The latter is constructed
by representing each $S=1$ spin as two separate $S=1/2$ spins,
binding pairs of these into antiferromagnetic dimers for each
exchange bond, and projecting the resulting state back onto the
subspace where $S_i^2=2$ on each site. This wave function is the
exact ground state of the Affleck-Kennedy-Lieb-Tasaki (AKLT)
model,\cite{Affleck87} and is schematically shown in the left
inset of Fig.~\ref{phase}. Each exchange link carries exactly one
valence bond, and the periodicity of the underlying crystal
lattice remains intact. Considerably less attention has been given
to a {\it different} quantum spin liquid ground state that is
realized in integral spin chains with alternating exchange
interactions. As the alternation parameter
$\delta=(J_1-J_2)/(J_1+J_2)$ deviates from zero (uniform chain),
the energy gap $\Delta$ decreases and closes at some critical
value $|\delta|=\delta_c\approx
0.26$,\cite{Affleck87-2,Yamamoto94,Yamamoto95,Yamamoto95-2,Yamamoto97,Kitazawa96,Kohno98}
 as illustrated in Fig.~\ref{phase}. Further increasing $|\delta|$ beyond this
quantum-critical point re-opens the spin gap. The ground state is
then no longer the Haldane state, but, instead, a dimerized one.
The corresponding valence bond wave function is shown in right
inset in Fig.~\ref{phase}. It contains two valence bonds on each
strong link and none at all on the weaker ones.

The two gapped quantum phases differ by their ``hidden''
symmetries.\cite{Kennedy92,Yamamoto97} The Haldane state is
characterized by a non-vanishing expectation value for the
antiferromagnetic string order parameter,\cite{Nijs89} related to
a breaking of a non-local $Z_2\times Z_2$
symmetry.\cite{Kennedy92} This order parameter vanishes in the
dimerized phase, where $Z_2\times Z_2$ remains completely intact.\cite{Kennedy92,Yamamoto97} The highly non-local
multi-spin correlation function that defines antiferromagnetic
strings  can not be expressed through the usual pair spin
correlation functions $\langle S_i^{(\alpha)}(0)\langle
S_j^{(\alpha)}(t)\rangle$. As a result, the ``hidden'' string
order can not be directly observed in scattering or other type of
experiments. In fact, one does not expect any {\it qualitative}
differences in observable spin correlation functions of
alternating $S=1$ chains with similar gap energies on different
sides of the phase diagram. Distinguishing the two phases in a
real $S=1$ alternating-chain compound is therefore a challenging
task, involving a careful quantitative analysis of the data.

A model alternating $S=1$-chain material suitable for experimental
studies was discovered only recently.\cite{Narumi2001} This
compound is \NTENP, NTENP for short, is structurally similar to
well-known Haldane-gap systems  NENP\cite{Regnault94+} and
NDMAP.\cite{Honda98,Zheludev2003+} Unlike the latter, NTENP
features a distinctive alternation of short and long bonds in the
antiferromagnetic $S=1$ Ni$^{2+}$ chains.\cite{Escuer97} Extensive
bulk measurements on NTENP were performed by Narumi et al. and are
reported in Ref.~\onlinecite{Narumi2001} Susceptibility
data\cite{Escuer97,Narumi2001} shows that in NTENP
$\Delta/J\approx 0.4$, almost the same as in a Haldane spin
chain.\cite{defJ} Anisotropy effects aside, on the simplified
phase diagram of an alternating $S=1$ chain\cite{Yamamoto95} shown
in Fig.~\ref{phase} NTENP must be located somewhere on the dashed
horizontal line. This line crosses the theoretical curve for
$\Delta(\delta)$ twice: at $\delta <\delta_c$ (near the Haldane
point $\delta=0$) and at $\delta\approx 0.40 >\delta_c$. Bulk
measurements can not directly probe the {\it microscopic}
alternation parameter $\delta$. Nevertheless, Hagiwara et al. were
able to conclude that NTENP is in the dimerized phase based on
indirect evidence, namely the behavior of non-magnetic impurities
in this material.\cite{Narumi2001} The main purpose of the present
work is use a microscopic probe (inelastic neutron scattering) to
{\it directly} measure exchange alternation and other crucial
parameters of {\it undoped} NTENP. Our experimental findings are
discussed in comparison with first-principles DMRG calculations.

\section{Experimental Approach}

Before describing the actual neutron scattering experiments
performed as part of this study, we shall discuss the measurement
strategies. In particular, we need to identify those measurable
physical quantities that are most sensitive to the effect of bond
alternation.

\subsection{Structural consideration} 
\begin{figure}
 \includegraphics[width=3.3in]{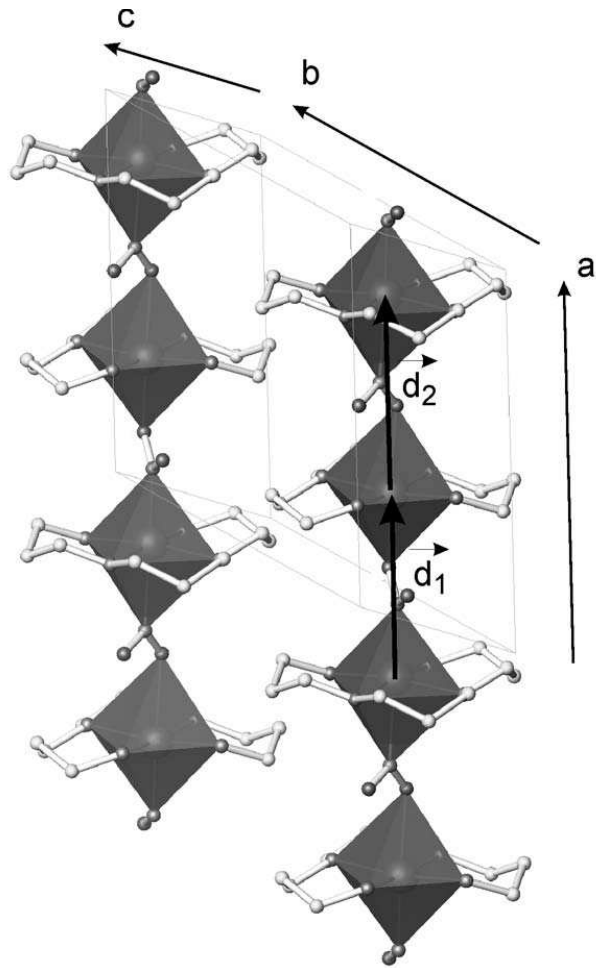}
 \caption{\label{struc} A schematic view of the bond-alternating $S=1$ Ni$^{2+}$ chains
 in the triclinic crystal structure of NTENP. The equatorial vertices of the Ni$^{2+}$ coordination
 octahedra are nitrogen atoms. The octahedra are coupled via chemically disordered NO$_2$ groups.
 Oxygen and nitrogen atoms are shown as dark grey spheres. Light grey spheres are carbon atoms.
 Hydrogen atoms and intercalated ClO$_4$ solvent molecules are not shown. The alternation of short and
 long bonds in the chains is characterized by bond vectors $\mathbf{d}_1$ and
 $\mathbf{d}_2$, respectively.}
\end{figure}

The triclinic crystal
structure of NTENP is visualized in
Fig.~\ref{struc}.\cite{Narumi2001,Hagiwara-private} The $S=1$
chains are composed of Ni$^{2+}$ ions octahedrally coordinated in
an organic environment. The chains run along the $a$ axis, one
chain per unit cell. While all Ni$^{2+}$ sites are
crystallographically equivalent, the Ni-Ni distances within the
chains alternate between $d_1=4.28$~\AA\ and $d_2=4.86$~\AA.
Intra-chain Ni-Ni links are covalent and pass trough structurally
disordered NO$_2$ groups. Inter-chain interactions are of Van der
Vaals nature and therefore much weaker. The crystallographic
symmetry is low, space group $P\overline{1}$. The lattice
constants at room temperature are $a=10.75$~\AA, $b=9.41$~\AA,
$c=8.79$~\AA, $\alpha=95.52^{\circ}$, $\beta=108.98^{\circ}$, and
$\gamma=106.83^{\circ}$.\cite{Escuer97} In the following
discussion we will use the following coordinate system: $x$ is
chosen along the $a^{\ast}$ axis, $z$ is along $c$, and $y$
completes a right-handed set of axes.

\subsection{Model Hamiltonian and observable energy scales} A model
Hamiltionian in for NTENP was proposed in
Ref.~\onlinecite{Narumi2001}. In the present work we shall employ
a slightly different, but equivalent form:\cite{defJ}
\begin{widetext}
\begin{equation}
\hat{H}=J(1-\delta)\sum_j\hat{\mathbf{S}}_{2j}\hat{\mathbf{S}}_{2j+1}+J(1+\delta)\sum_j\hat{\mathbf{S}}_{2j}\hat{\mathbf{S}}_{2j-1}+D\sum_j(S_j^{(x)})^2
\label{ham}
\end{equation}
\end{widetext}
Inter-chain spin interactions are expected to be very weak and are
not included in the above expression. The exchange constant $J$
is, itself,  not directly observable experimentally. However, it
can be reliably inferred from an analysis of the measured
temperature dependence of bulk susceptibility: for NTENP $J\approx
3.4$~meV.\cite{Escuer97,Narumi2001}

A directly measurable quantity related to $J$ is zone-boundary
energy $\hbar\omega_{\mathrm{zb}}$, defined as the minimal energy
of magnetic excitations with momentum transfer
$q_{\mathrm{zb}}=\frac{\pi}{a}$, where $a$ is the distance between
next-nearest-neighbor spins. In NTENP $a$ is simply the lattice
constant. For a uniform (Haldane) chain with $\delta=0$ numerical
simulations show that $\hbar\omega_{\mathrm{zb}}\approx
2.7J$.\cite{Golinelli92-2,Takahashi93,Sorensen94,Yamamoto95-3} It
is easy to verify that $\hbar\omega_{\mathrm{zb}}=2J$ for isolated
dimers ($\delta=1$). The complete $\delta$-dependence of
$\hbar\omega_{\mathrm{zb}}$ was recently determined in a
systematic numerical study.\cite{Suzuki2003} It is important to
note that unlike $\Delta/J$, $\hbar\omega_{\mathrm{zb}}/J$ is a
{\it monotonic} function of $\delta$, and can be used to {\it
unambiguously} determine whether a particular material in in the
dimerized or Haldane phases.

Another directly observable energy scale is spin wave velocity
$v$. Numerically, $v\approx 2.49 J$ for a uniform
chain\cite{Sorensen94} and $v\equiv 0$ for the other limiting case
of isolated $S=1$ dimers. The gap energies $\Delta_\alpha$ for
different spin polarizations are also experimentally accessible.
For an isotropic uniform chain $\Delta\approx 0.41
J$.\cite{Golinelli92-2,Takahashi93,Sorensen94,Yamamoto95-3} For
isolated dimers with $\delta=1$  $\Delta=2J$. According to a
simple perturbation theory argument,\cite{Regnault93,Golinelli93}
the polarization-averaged energy gap $\overline{\Delta}\equiv
\frac{1}{3}\sum_\alpha\Delta_\alpha$ is, to a good approximation,
the same as in the isotropic system with $D=0$. The gap for
excitations polarized along the $x$, $y$ and $z$ axes can then be
written as:
\begin{eqnarray}
\Delta_x &=&\overline{\Delta}+2\tilde{D},\nonumber \\
\Delta_z= \Delta_y &=& \overline{\Delta}-\tilde{D}.
\end{eqnarray}
The observable splitting $\tilde{D}$ is proportional to the
microscopic anisotropy parameter $D$ in the Hamiltonian. Numerical
simulations indicate that for a uniform chain $\tilde{D}\approx
\frac{2}{3}D$.\cite{Golinelli92,Golinelli93} For isolated dimers
one simply has $\tilde{D}=D$. For NTENP the gap energies can be
estimated from high-field magnetization measurements of Narumi
{\it et al}.\cite{Narumi2001} The critical field at which the gap
for one of the spin polarizations is driven to zero by the Zeeman
effect is given by
$g^{\alpha}\mu_{\mathrm{B}}H_{c}^{(\alpha)}=\sqrt{\Delta_\beta
\Delta_\gamma}$.\cite{Tsvelik90,Golinelli92,Golinelli93-2} For
NTENP $H_c^{(x)}=9.3$~T, $H_c^{(y)}\approx H_c^{(x)}=12.4$~T, and
$g=2.14$. This gives $\Delta_y\approx\Delta_z=1.15$~meV and
$\Delta_x=2.06$~meV. From this one gets
$\overline{\Delta}=1.45$~meV ($\overline{\Delta}/J\approx 0.43$)
and $\tilde{D}\approx 0.3$~meV ($\tilde{D}/J\approx
0.1$).\footnote{Note the discrepancy with
Ref.~\protect\onlinecite{Narumi2001}, where the authors seemingly
fail to distinguish between $D$ and $\tilde{D}$. In any case,
their quoted estimate ($D/J=0.35$ using the symmetric definition
of $J$) is hard to reconcile with the very small anisotropy of
critical field and gap energies that the authors report and that
our neutron scattering experiments confirm. We believe that the
discrepancy is due to inconsistent notations or a trivial mistake
in Ref.~\protect\onlinecite{Narumi2001}.}

The important energy scales $\overline{\Delta}$,
$\hbar\omega_{\mathrm{zb}}$, $v$ and $\tilde{D}$ can be
straightforwardly measured in inelastic neutron scattering
experiments, by mapping out the dispersion relation of magnetic
excitations. They do not, however, carry any {\it direct}
information on the level of dimerization in the system.

\subsection{Exploiting the 1st-moment sum rule}

Additional insight can be drawn from an analysis of neutron
scattering {\it intensities} of magnetic excitations. In fact,
these intensities directly relate to the strengths of individual
magnetic bonds. One way of extracting this information is by
making use of the Hohenberg-Brinkman first moment sum rule for the
magnetic dynamic structure
factor.\cite{Hohenberg74,Zaliznyak-sumrules} This method has been
previously successfully applied to the analysis of inelastic
neutron scattering data on several occasions: for recent examples
see Refs.~\onlinecite{Xu2000,Zaliznyak2001}. For the Hamiltonian
(\ref{ham}) the sum rule, an {\it exact} expression, can be
written as:\cite{Zaliznyak-sumrules}
\begin{eqnarray}
\int_{-\infty}^{\infty}(\hbar\omega)\mathcal{S}^{\alpha\alpha}(\mathbf{q},\omega)d(\hbar
\omega) =-\frac{1}{2N}\langle \left[\hat{S}^{(\alpha)}_{\mathbf{q}}, \left[\hat{S}^{(\alpha)}_{\mathbf{-q}}, \hat{H}\right]\right]\rangle= \nonumber\\
 -\sum_\beta 2 J_1\sin^2(\mathbf{q}\mathbf{d}_1/2)(1-\delta_{\alpha\beta})\langle
 S_{2j}^{(\beta)} S_{2j+1}^{(\beta)}\rangle-\nonumber\\
 -\sum_\beta 2 J_2\sin^2(\mathbf{q}\mathbf{d}_2/2)(1-\delta_{\alpha\beta})\langle
 S_{2j}^{(\beta)} S_{2j-1}^{(\beta)}\rangle-\nonumber\\
 -2D(1-\delta_{x\beta})\left[2\langle(S_j^{(x)})^2\rangle+\langle(S_j^{(\alpha)})^2\rangle-S(S+1)\right].
 \label{SR1}
 \end{eqnarray}

 Here $\mathbf{d}_1$ and $\mathbf{d}_2$
are real-space vectors chosen along the short and long bonds in
the chains, respectively, and $\alpha$ and $\beta$ label the
coordinate axes: $x$, $y$ and $z$. In practice it may be quite
difficult to separately measure all three diagonal components of
$\mathcal{S}^{\alpha\alpha}$. Fortunately, for NTENP the ratio
$D/J$ is only about 10\% (see discussion above), and, to a good
approximation, the correlators $\langle
 S_{2j}^{(\beta)} S_{2j+1}^{(\beta)}\rangle$ and $\langle
 S_{2j}^{(\beta)} S_{2j-1}^{(\beta)}\rangle$ are independent of the
 subscript $\beta$.
The last term in Eq.~\ref{SR1} scales as $(D/J)^2$
(Ref.~\onlinecite{Regnault93}) and can be entirely neglected in
our case. Under these assumptions the sum rule becomes:
\begin{eqnarray}
\int_{-\infty}^{\infty}(\hbar\omega)\mathcal{S}^{\alpha\alpha}(\mathbf{q},\omega)d(\hbar
\omega) =\nonumber\\
 -\frac{4}{3}E_1\sin^2(\mathbf{q}\mathbf{d}_1/2)
 -\frac{4}{3}E_2\sin^2(\mathbf{q}\mathbf{d}_2/2).
 \label{SR2}
\end{eqnarray}
The quantities $E_1=J_1\langle
\mathbf{S}_{2j}\mathbf{S}_{2j+1}\rangle$ and $E_2=J_2\langle
\mathbf{S}_{2j}\mathbf{S}_{2j-1}\rangle$ are ground state {\it
exchange energies} associated with the strong and weak bonds,
respectively. Due to the translational invariance, they do not
depend on the site index $j$.

Eq.~\ref{SR2} directly relates the intensities measured in an
inelastic neutron scattering experiment to the {modulation of
exchange energy} in the spin chains
\begin{equation}
\tilde{\delta}=\frac{E_1-E_2}{E_1+E_2}. \label{ratio}
\end{equation}
While certainly {\it not} equivalent to $\delta$, $\tilde{\delta}$
is a very natural measure of the magnitude of ``dimerization'' of
the ground state. In the isotropic (Heisenberg) case of $D=0$ it
can be directly expressed through $\delta$ and the ground state
energy $E(\delta)\equiv \langle H \rangle$:

\begin{eqnarray}
\tilde\delta &=& \Big\langle\sum_j (1+\delta)S_{2j}S_{2j+1}
                       -\sum_j (1-\delta)S_{2j-1}S_{2j}\Big\rangle
                       / E(\delta)\nonumber\\
&=& \delta + (1-\delta^2) {\partial\over\partial\delta}
\ln{E(\delta)}. \label{deltadelta}
\end{eqnarray}

For practical applications Eq.~\ref{SR2} can be further simplified
if one assume the {\it single mode approximation}
(SMA):\cite{Auerbach}
\begin{equation}
 \mathcal{S}^{\alpha\alpha}(\mathbf{q},\omega)\approx\mathcal{S}^{\alpha\alpha}(\mathbf{q})
\delta(\omega-\omega_{\mathbf{q},\alpha}).
 \label{SMA}
\end{equation}
The sum rule for a bond-alternating chain is then written as:
\begin{equation}
\mathcal{S}^{\alpha\alpha}(\mathbf{q}) \approx
-\frac{4}{3\omega_\mathbf{q}}\left[
 E_1\sin^2(\mathbf{q}\mathbf{d}_1/2)
 +E_2\sin^2(\mathbf{q}\mathbf{d}_2/2)\right].
 \label{SR3}
\end{equation}
For isolated dimers expression \ref{SMA} is exact. For a uniform
spin chain the SMA works remarkably well in most of the Brillouin
zone, especially in the vicinity of the 1D AF zone-center where
the Haldane gap is observed.\cite{Ma92,Xu96,Zaliznyak2001} Near
the quantum critical point, where $\Delta$ vanishes, the SMA will
fail entirely. However, for NTENP $\Delta/J$ is similar to that in
a uniform chain, and the SMA should still be reliable near the 1D
zone-center $\mathbf{q}_0$,
$\mathbf{q}_0(\mathbf{d}_1+\mathbf{d}_2)=2\pi$. In this range the
SMA dispersion relation can be written in the standard
``relativistic'' form:
\begin{equation}
 (\hbar \omega_{\mathbf{q},\alpha})^2=\Delta_\alpha^2+v^2\sin^2(\mathbf{qa})
 \label{disp}
\end{equation}

\subsection{Application to NTENP} 
\begin{figure}
 \includegraphics[width=3.3in]{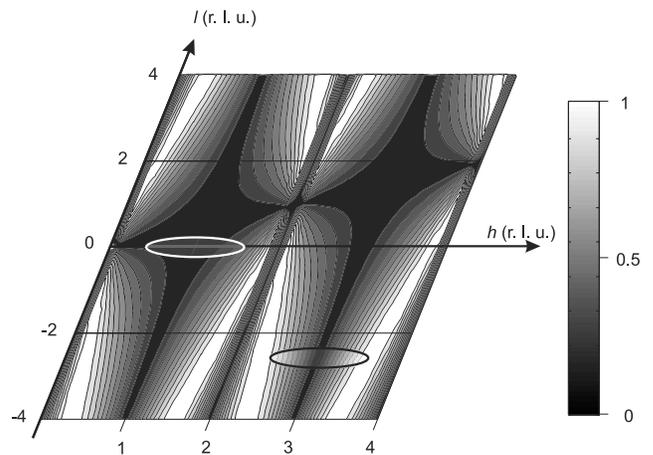}
 \caption{\label{contrast} Contrast ration $C(\mathbf{q})$ (Eq.~\protect\ref{Ccontrast}) for distinguishing
 ground state exchange energies based on inelastic neutron
 intensities as a function of momentum transfer in the $(h,0,l)$
 reciprocal-space plane of NTENP. The circled areas are regions of reciprocal space
 where most of the inelastic data were measured.}
\end{figure}

The main experimental
difficulty in using Eq.~\ref{SR3} to estimate the ratio
$\tilde{\delta}$ in NTENP is the fact that the bond vectors
$\mathbf{d}_1=0.521\mathbf{a}+0.0246\mathbf{b}-0.0424\mathbf{c}$
and
$\mathbf{d}_2=0.479\mathbf{a}-0.0246\mathbf{b}+0.0424\mathbf{c}$
are quite close in this material. Fig.~\ref{contrast} shows a
grayscale and contour plot of the effective ``contrast'' ratio
\begin{equation}
C(\mathbf{q})\equiv\frac{\sin^2(\mathbf{q}\mathbf{d}_1/2)-\sin^2(\mathbf{q}\mathbf{d}_2/2)}{\sin^2(\mathbf{q}\mathbf{d}_1/2)+\sin^2(\mathbf{q}\mathbf{d}_2/2)}\nonumber
\label{Ccontrast}
\end{equation}
as a function of momentum transfer in the $(h,0,l)$
reciprocal-space plane in NTENP. This ratio is a good measure of
our sensitivity to $\tilde{\delta}$. Immediately one can see that
$E_1$ and $E_2$ can not be distinguished based on the data
collected at the 1D AF zone-centers, where $C(\mathbf{q})$
vanishes. This is rather unfortunate, since it is at these wave
vectors that the dispersion $\omega_\mathbf{q}$ is a minimum, and
excitation intensity is maximized due to the $1/\omega_\mathbf{q}$
factor in Eq.~\ref{SR3}. A high contrast is achieved away from the
1D AF zone-centers, and at large momentum transfers. However,
under these conditions the magnetic scattering is weakened by the
$1/\omega_\mathbf{q}$ coefficient and the effect of ionic magnetic
form factors. Away from the AF zone-centers the applicability of
the SMA also becomes questionable. Finally, the phonon background
becomes progressively important at large $|q|$ and interferes with
the measurements. In our experiments we have found that a
reasonable compromise between intensity, contrast and noise level
can be achieved on either side of the $h=3$ 1D zone-center. Most
of the data described in Section~\ref{global} below were collected
in that region of reciprocal space, represented in
Fig.~\ref{contrast} by the circled area in the lower-right.

\section{Experimental Procedures}
Translucent dark-purple plate-like single crystal samples of 50\%
deuterated NTENP  were grown in aqueous solution. Four such
crystals were co-aligned in one ``supersample'' with a total mass
of 1.5~g and mosaic spread of 2.5$^\circ$. Inelastic neutron
scattering experiments were performed at two different facilities.
Lower-energy excitations (up to 3~meV energy transfer) were
investigated using the SPINS cold-neutron 3-axis spectrometer
installed at the NIST Center for Neutron Research. Neutrons with a
fixed final energy $E_f=3.7$~meV were utilized with  a BeO filter
after the sample. A Pyrolitic Graphite PG(002) monochronmator was
used in combination with a flat (Setup I) or horizontally focusing
(Setup II) PG analyzers. Additional  beam collimation was provided
by the neutron guide and $\mathrm{(open)}-80'
-80'-\mathrm{(open)}$ collimators (Setup II employed a radial
post-sample collimator). Thermal-neutron studies were performed
using the HB-1 3-axis spectrometer installed at the High Flux
Isotope Reactor, Oak Ridge National Laboratory. The data were
collected with $E_i=13.5$~meV fixed-incident energy neutrons
(Setup III). PG(002) reflections were employed in both
monochromator and analyzer. A PG filter was installed in front of
the sample to eliminate higher-order beam contamination. Beam
collimation was $48'-40'-40'-240'$.

The sample being only partially deuterated led to a substantial
geometry-dependent attenuation of the neutron beam due to
incoherent scattering from hydrogen nuclei. This effect is
equivalent to neutron absorption, and can be fully compensated for
using the technique described in Ref.~\onlinecite{Zheludev2000}.
For every inelastic scan measured, one performs a separate elastic
scan to determine the effective neutron transmission corrections.
For every point of the ``transmission'' scan the sample rotation
and scattering angles are set exactly as in the inelastic scan.
 Unless a Bragg condition is accidentally satisfied in the
sample, the main contribution to scattering in the
``transmission'' scan is due to incoherent elastic processes in
the sample. To a good approximation, the corresponding cross
section is isotropic and independent of neutron energy. The
intensity detected in the ``transmission'' scan is therefore
directly proportional to the neutron transmission in the sample.
Normalizing the original inelastic scan by the measured
transmission correction not only gets rid of absorption effects,
but also compensates for the any geometric corrections that occur
when a large asymmetric sample rotates in a finite-size neutron
beam in the course of the scan. In various scans measured in the
present work the effective transmission coefficient varied by
roughly a factor of 2 in the course of each scan or between
different scans.

\begin{figure}
 \includegraphics[width=3.3in]{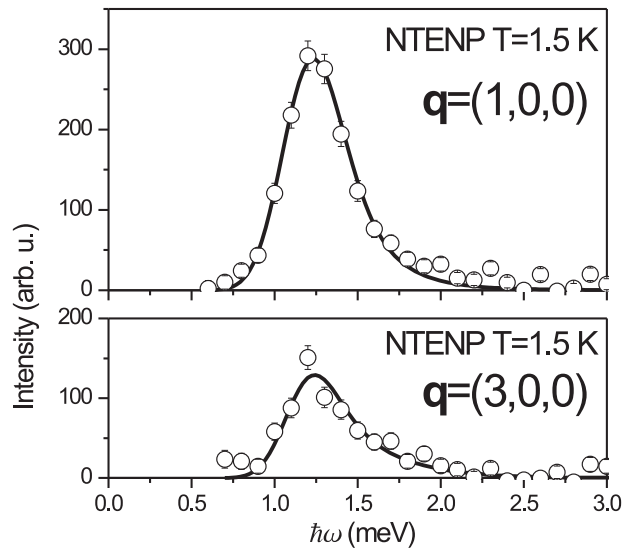}
 \caption{\label{constq} Inelastic scans collected at the $(1,0,0)$
 and $(3,0,0)$ 1D antiferromagnetic zone-centers in NTENP (symbols). The data were
 taken  using a cold neutron 3-axis spectrometer with a flat analyzer. The solid lines
 are fits to the data as described in the text.}
\end{figure}

\begin{figure}
 \includegraphics[width=3.3in]{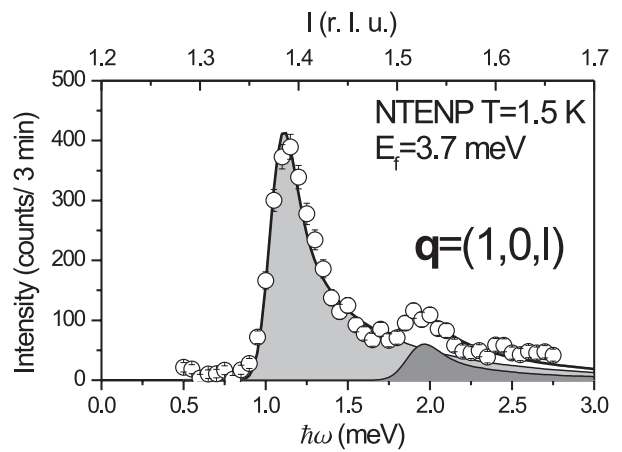}
 \caption{\label{focus} Inelastic scan at the $\mathbf{q}=(1,0,l)$ 1D antiferromagnetic zone-center measured using
 a horizontally-focusing analyzer. The heavy solid line is a fit to the data as described in the
 text. The light-grey and dark-grey shaded areas represent partial contributions of gap
 excitations polarized parallel and perpendicular to the crystallographic $(b,c)$ plane, respectively.}
\end{figure}

\section{Experimental results}
\subsection {Gap energies: constant-$q$ data}
The gap energies and anisotropy splitting of the excitation
triplet were accurately measured using cold-neutron Setups I and
II. Energy scans collected at 1D AF zone-centers $(1,0,0)$
(Fig.~\ref{constq}a) and $(3,0,0)$ (Fig.~\ref{constq}b) with Setup
I show a single sharp peak at about 1.2~meV energy transfer. The
background for these scans was measured, point-by-point at
$\mathbf{q}=(1.5,0,0)$ and $(2.5,0,0)$, respectively. As shown,
the scans are corrected for transmission effects. The data in
Fig.~\ref{constq} were collected with the scattering vector
$\mathbf{q}$ directed perpendicular to the $(b,c)$
crystallographic plane, and therefore represent fluctuations of
$y$ and $z$ spin components of the triplet. The resolution of the
present experiment is insufficient to unambiguously resolve the
gaps for $y$- and $z$-polarized modes.
 To detect the $x$-axis spin fluctuations
we performed additional measurements with a large momentum
transfer perpendicular to the chains. In the focusing-analyzer
mode (Setup II) the scattering vector was at all times maintained
on the ($1,0,l)$ reciprocal-space rod. For each energy transfer,
the transverse momentum transfer $l$ was chosen to have the sample
chain axis parallel to the scattered neutron beam. This geometry
optimizes $q$-resolution along the chains.  The resulting scan is
shown in Fig.~\ref{focus}. Due to the intrinsic
polarization-dependence of the magnetic scattering cross-section,
the largest contribution to this scan is from $y$ and
$x$-polarized excitations. The new feature observed at about
1.8~meV energy transfer was thus attributed to $x$-axis spin
fluctuations. Note that the lower-energy peak in Fig.~\ref{focus}
appears at a slightly lower energy than in the scan taken with the
scattering vector parallel to the chain axis. This behavior is
likely due to a weak dispersion of excitations perpendicular to
the chains.

\begin{figure}
 \includegraphics[width=3.3in]{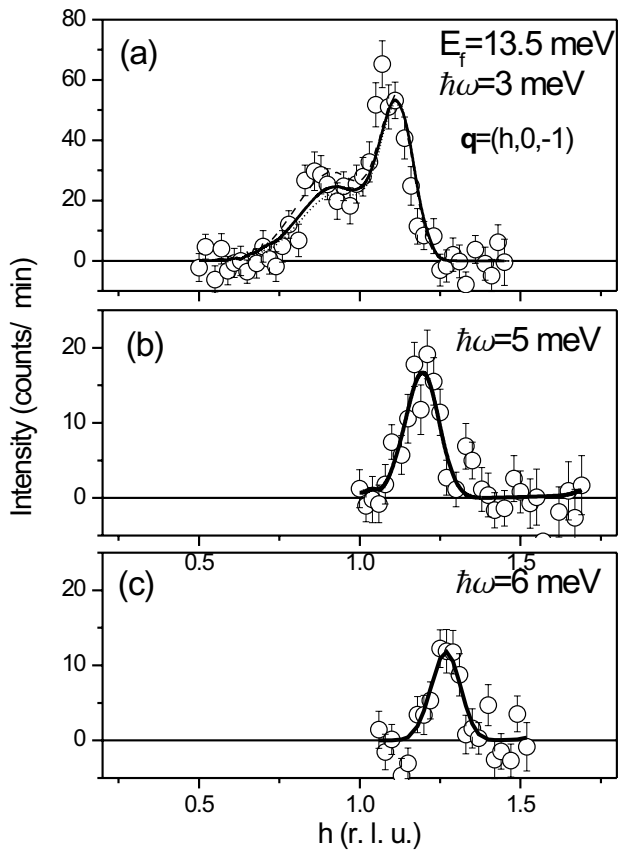}
 \caption{\label{noeffect} Constant-$E$ scans measured in NTENP in the vicinity of the for
 $(1,0,0)$AF zone-center (symbols). The dashed and dotted lines in
 (a)  are profiles simulated assuming a cross section function as
 given by Eqs.~\protect\ref{SR3}--\protect\ref{disp} with
 $\tilde{\delta}=-1$ and $\tilde{\delta}=1$, respectively. The
 sold line in (a) is a global fit to several scans, as described
 in the text. The solid lines in (b) and (c) are fits to single
 scans.}
\end{figure}

\begin{figure}
 \includegraphics[width=3.3in]{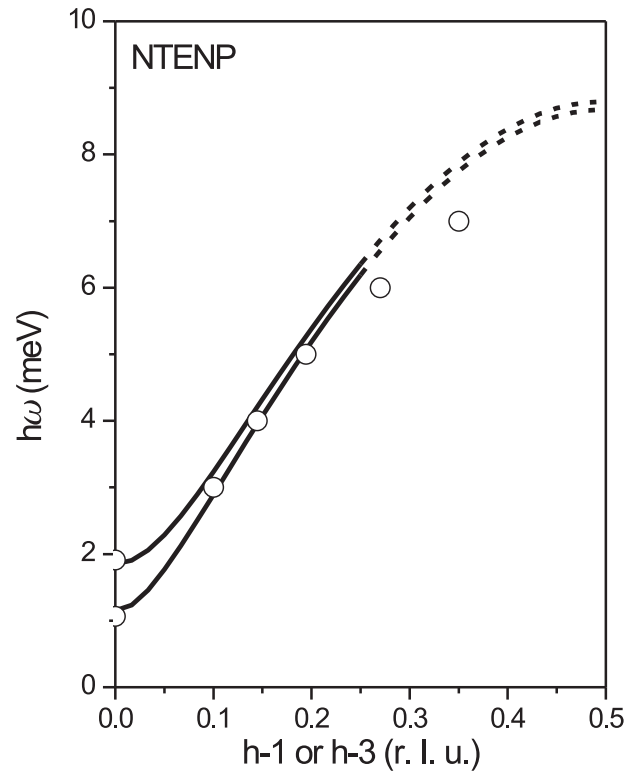}
 \caption{\label{dispfig} Measured dispersion relation of magnetic excitations in NTENP (symbols).
 Lines are the result of a global fit of a model cross section to the neutron data as described
 in the text.}
\end{figure}

The constant-$Q$ scans were analyzed assuming a simple single-mode
cross section as given by Eqs.~\ref{SMA}, \ref{SR3} and
\ref{disp}. The spin wave velocity was fixed at $v=8.6$~meV, as
separately determined from the analysis of constant-$E$ scans
described below. The usual polarization factors for unpolarized
neutrons determined the relative intensities of the in-plane and
out-of-plane  spin fluctuations. In addition, a magnetic form
factor for Ni$^{2+}$ was explicitly included in the cross section
function. The resulting model for the dynamic structure factor was
numerically convoluted with the spectrometer resolution function,
calculated in the Cooper-Nathans approximation. The only
adjustable parameters of the model were the gap energies
$\Delta_x$, $\Delta_y$ and $\Delta_z$ and an overall intensity
scaling factor. A very good fit to the scans with zero transverse
momentum transfer was obtained with
$\Delta_y=\Delta_z=1.16(0.01)$, as is shown in solid lines in
Fig.~\ref{constq}. Slightly smaller energies of the doublet were
obtained by analyzing the data shown in Fig.~\ref{focus}. Here the
gap energies were found to be $\Delta_y=\Delta_z=1.07(0.01)$ and
$\Delta_x=1.91(0.02)$~meV. These values correspond to
$\overline{\Delta}=1.35(2)$~meV and $\tilde{D}=0.28(2)$~meV, in
good agreement with bulk measurements of Narumi {\it et
al}.\cite{Narumi2001}

\begin{figure}
 \includegraphics[width=3.3in]{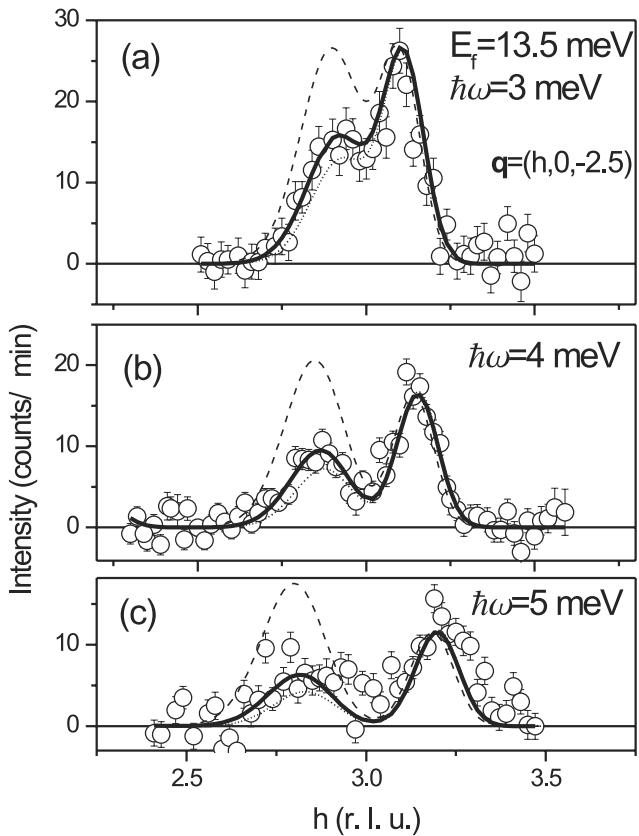}
 \caption{\label{effect} Constant-$E$ scans measured in NTENP using a thermal-neutron
 setup in the vicinity of the
 $(3,0,-2.5)$ AF zone-center (symbols). All lines are as in
 Fig.~\protect\ref{noeffect}(a).}
\end{figure}

\subsection {Dispersion relation: constant-$E$ data}
The dispersion relation for single-mode excitations in NTENP was
measured in constant-$E$ scans. In order to optimize magnetic
intensities (form factor) these data were collected around the
$(1,0,0)$ AF zone-center (left circled area in
Fig.~\ref{contrast}). Typical scans are plotted in open symbols in
Fig.~\ref{noeffect}. To take into account resolution effects, but
to avoid constraining the dispersion relation to the postulated
sinusoidal form, each scan was {\it separately} analyzed using the
model cross section described above. The adjustable parameters for
each scan were an intensity prefactor and the spin wave velocity
$v$. The gap energies were fixed at the values determined using
cold neutrons (see above). Typical fits are shown in
Fig.~\ref{noeffect}(b) and (c) in solid lines. The dispersion
relation deduced from such fits to individual scans is plotted in
open symbols in Fig.~\ref{dispfig}.

\begin{figure}
 \includegraphics[width=3.3in]{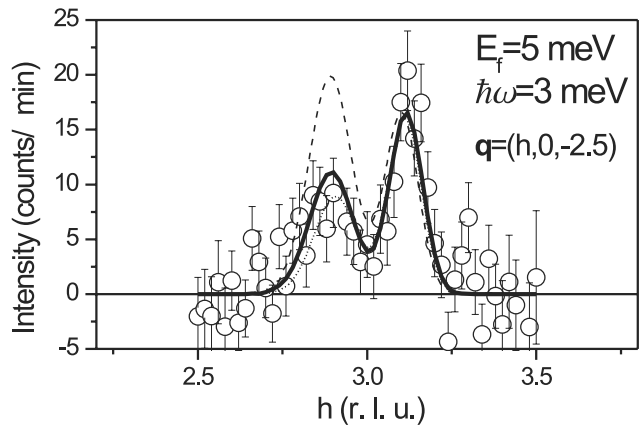}
 \caption{\label{spinsconste} Constant-$E$ scan measured in NTENP using a cold-neutron
 setup in the vicinity of the
 $(3,0,-2.5)$ AF zone-center (symbols). All lines are as in
 Fig.~\protect\ref{noeffect}(a).}
\end{figure}

\subsection {Exchange energy modulation and global fits}
\label{global} All scans described above are fairly insensitive to
the distribution of exchange energies in the chains, due to a
small contrast ratio at the particular wave vectors. To illustrate
this, in Fig.~\ref{noeffect}(a) we have plotted peak profiles
simulated using Eqs.~\ref{SR3}--\ref{disp}, and assuming $E_1=0$
(dashed line) or $E_2=0$ (dotted line). To within the accuracy of
our measurements, the two profiles are almost identical. A much
better contrast (although a significantly smaller intensity) was
achieved in constant-$E$ scans collected along the $(h,0,-2.5)$
reciprocal-space rods around $h=3$ (right circled area in
Fig.~\ref{contrast}). These data are plotted in symbols in
Fig.~\ref{effect}. A similar constant-$E$ scan was measured using
cold neutrons and Setup I, and is plotted if
Fig.~\ref{spinsconste}. In both figures the dashed and dotted
lines are  simulations for $E_2=0$ or $E_1=0$, respectively,
assuming $v=8.6$~meV. From these data the strong dimerization in
NTENP becomes apparent: the ground state exchange energy is
primarily concentrated on the shorter bonds.

To quantify this observation we performed a {\it global} fit to
the data collected in constant-$E$ scans measured for energy
transfers up to $5$~meV with setups I and III. The adjustable
parameters were two intensity scaling factors  (one for each
setup), the spin wave velocity $v$ and the bond-alternation
parameter $\tilde{\delta}$. Very good fits are obtained with
$v=8.6(1)$~meV and $\tilde{\delta}=0.42 (+0.2,-0.1)$. The large
asymmetric error bar on $\tilde{\delta}$ is unavoidable due to the
technical difficulties associated with low intensity, small
contrast ratio and transmission corrections.

The simple sinusoidal dispersion curves postulated in
Eq.~\ref{disp} is plotted in lines in Fig.~\ref{dispfig} using the
experimentally determined gap energies and spin wave velocity. At
higher energies the experimental data points clearly lie below
these curves. While it is difficult to extrapolate the
experimental result to the zone-boundary, one can roughly estimate
$\hbar \omega_{\mathrm{ZB}}$ to be smaller than $v$ by about
1~meV.

All physical parameters obtained for NTENP in our neutron
scattering studies are summarized in Table~\ref{params} in
comparison with those obtained by Narumi {\it et al.} using bulk
techniques\cite{Narumi2001} and to known exact results for uniform
and fully dimerized isotropic $S=1$ chains.

 \begin{table*}
 \caption{\label{params} Physical parameters for the alternating $S=1$ chains in NTENP in comparison
 with those for uniform and fully dimerized isotropic $S=1$ spin chains.}
 \begin{ruledtabular}
 \begin{tabular}{l l l l l}
 & NTENP\footnote{Bulk measurements: Ref.~\protect\onlinecite{Escuer97,Narumi2001}} & NTENP\footnote{Neutron scattering: this work, assuming $J=3.4$~meV.}
 & Uniform chain & Isolated dimers\\
 \hline\\
 $J$ & 3.4~meV & --& --& --\\
 $\overline{\Delta}$ & $1.45$~meV $\approx 0.42 J$ & $1.35(2)$~meV $\approx 0.40(1) J$& $0.41 J$ & $2 J$\\
 $\tilde{D}$ & $0.3$~meV $\approx 0.1 J$ & $0.28(2)$~meV $\approx 0.083(6) J$& $\frac{2}{3}D$\protect\cite{Golinelli92,Golinelli93} & $D$\\
 $v$ & -- & $8.6(1)$~meV$\approx 2.5(1) J$ & $2.5 J$\protect\cite{Sorensen94} & 0\\
 $\hbar\omega_{\mathrm{ZB}}$ & -- & $\approx 7.5(5)$~meV$=2.2(1) J$& $2.7 J$\protect\cite{Sorensen94} & $2J$\\
 $\tilde{\delta}\equiv \frac{E_1-E_2}{E_1+E_2}$ & --& 0.42& 0 & 1\\
 \hline
 $\delta\equiv \frac{J_1-J_2}{J_1+J_2}$ & $0.40$ or $\approx 0$ \footnote{From $\overline{\Delta}/J$ with $\overline{\Delta}$ deduced from high-field magnetization data  and numerical results of Ref.~\protect\onlinecite{Yamamoto95}.} & $0.37(1)$ or $0.06(2)$ \footnote{From $\overline{\Delta}/J$ with directly measured $\overline{\Delta}$ and numerical results of Ref.~\protect\onlinecite{Yamamoto95}.} & 0 & 1\\
 & & 0.30(0.05)\footnote{From $\hbar\omega_{\mathrm{ZB}}/J$ with directly measured  $\hbar\omega_{\mathrm{ZB}}$ and numerical results of Ref.~\protect\onlinecite{Suzuki2003}}\\
 & & 0.24(-0.04,+0.08)\footnote{From directly measured $\tilde{\delta}$ and numerical results of this work.}\\
 \end{tabular}
 \end{ruledtabular}
 \end{table*}

\section{Numerical calculations}
\begin{figure}
  \includegraphics[width=3.3in]{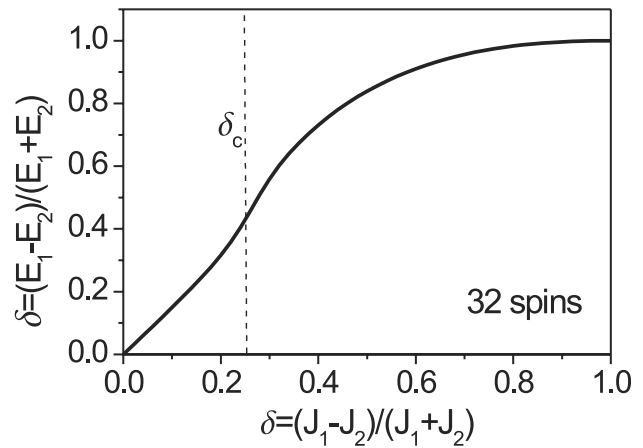}
 \caption{\label{dmrgfig} Alternation of exchange energies
$\tilde\delta=(E_1-E_2)/(E_1+E_2)$ as a function of the
alternation of exchange constants $\delta$, computed for a 32-site
bond-alternating $S=1$ chain with periodic boundary conditions.
The quantum phase transition at $\delta_c$ corresponds to an
inflection point on the calculated curve.}
\end{figure}

In order to relate the measured exchange energy modulation
parameter $\tilde{\delta}$ to the alternation of exchange
constants $\delta$, we performed a numerical study of the model
Hamiltonian (\ref{ham}), assuming a vanishing anisotropy $D=0$.
The ground state energy $E(\delta)$ was computed as a function of
$\delta$ using the density matrix renormalization group (DMRG)
method\cite{White92,White93,Peschel99} for a chain of 32 spins
with periodic boundary conditions. The parameter $\tilde{\delta}$
was then obtained using Eq.~\ref{deltadelta}. These results are
plotted in Fig.~\ref{dmrgfig}. The computed curve is monotonic and
quite smooth in the studied domain of parameter space. Crossing
the quantum-critical point at $\delta_c\approx 0.26$ is marked
only by a weak point-of-inflection-type anomaly. We recall that at
$\delta_c$ the correlation length diverges. However, for our
particular calculation, this does not appear to be a problem. The
observable (\ref{deltadelta}) is very smooth and is not subject to
significant finite-size corrections. It was verified that doubling
of the system size changed the results shown in Fig.~\ref{dmrgfig}
by less than one percent.

To justify the use of the isotropic model for computing
$\tilde{\delta}$ as a function of $\delta$, we also studied the
behavior of the expectation value of the anisotropy term in the
Hamiltonian \ref{ham}. It was found that that its contribution to
the ground state energy is practically independent of $\delta$ and
is equal to about $2/3 D$ per spin.

 \section{Discussion}
Our neutron results present new opportunities to determine the
magnitude of bond-alternation in NTENP and to place this material
on the phase diagram. First of all, $\delta$ can be estimated from
the gap energies. Using the numerical results of
Ref.~\onlinecite{Yamamoto95} and assuming
$J=$3.4~meV\cite{Narumi2001}, for $\overline{\Delta}=1.35(2)$~meV
one gets $\delta=0.37(1)$, assuming NTENP is on the dimerized side
of the phase diagram. The alternative is $\delta=0.06(2)$ in the
Haldane phase, and the distinction can not be made based on gap
measurements alone.

The numerical results of the previous section allow us to {\it
independently} estimate $\delta$ based the measured exchange
energy alternation. The experimental value
$\tilde\delta=0.42(+0.2,-0.1)$ corresponds to $\delta\approx
0.24(-0.04,+0.08)$. This value is within $1.5 \sigma$ of the
estimate based on the measured gap energy, with the assumption
that NTENP is in the dimerized phase. Assuming that NTENP was in
the Haldane phase with $\delta=0.06$ would imply
$\tilde{\delta}\approx 0.05$, in considerably worse agreement with
experiment. The error bars associated with our measurements of
$\tilde\delta$ are rather large, and the method itself relies on
the uncontrolled single-mode approximation. However, when combined
with gap measurements, these data cleraly confirm that NTENP has a
dimerized, rather than Haldane ground state. The same conclusion
can be reached by comparing the measured zone-boundary energy to
numerical results of Ref.~\onlinecite{Suzuki2003}, according to
which our experimental estimate
$\hbar\omega_{\mathrm{ZB}}/J\approx 2.2(1)$ corresponds,
unambiguously, to $\delta=0.3(0.05)$.

It is important to emphasize that our combination experimental
approach allows to {\it uniquely} determine $\delta$ and decide
whether the ground state is dimerized or not, based on measured
properties of the {\it undoped} material. This eliminates a
possible ambiguity associated with guessing the ground state from
the behavior of non-magnetic impurities, as was done in
Ref.~\onlinecite{Narumi2001}. The problem is that the ground
states of {\it chain-fragments} in the impurity-doped system are
different from the ground state of defect-free chains. In
particular, for the Haldane phase ($\delta<\delta_c$) there are
effective interactions between the two liberated $S=1/2$ spins on
the ends of each chain fragment.\cite{Hagiwara90,Glarum91,Mitra92}
Depending on the parity of the number of magnetic sites in the
fragment, the interaction is antiferro- or
ferromagnetic.\cite{Lieb62} In the former case, the fragment has a
{\it non-magnetic} $S=0$ ground state and does not produce any ESR
signal. In odd-length fragments, however, end-chain spins combine
to form an $S=1$ triplet. At sufficiently low temperature any ESR
experiment will observe $S=1$ (rather than $S=1/2$) degrees of
freedom, {\it just like in the dimerized phase, for
$\delta>\delta_c$}.

\section{Conclusion}
In summary, we have performed a combination of inelastic neutron
scattering measurements to determine the key characteristics of
the bond-alternating $S=1$ quantum spin chain compound NTENP. The
results unambiguously indicate a strong but incomplete
dimerization of the ground state in this material. Future studies,
perhaps using larger and fully deuterated single crystal samples,
will concentrate on features of the spectrum that are beyond the
simplified single-mode picture.

\acknowledgments We are profoundly grateful to Prof. M. Hagiwara
(RIKEN) for allowing us to use unpublished X-ray diffraction data
for NTENP and to Dr. L.-P. Regnault (CEA Grenoble) for sharing
some preliminary results of his independent experiments on this
material. One of the us (AZ) would like to thank Prof. Tchenyshev
(Johns Hopkins University) for explaining some finer aspects of
the theory of integral quantum spin chains, and Dr. I. Zaliznyak
for making his unpublished review of sum rules in magnetic neutron
scattering freely available. This work was supported in parts by
the U. S. Department of Energy under Contract Nos.
DE-AC05-00OR22725 (Oak Ridge National Laboratory) and
DE-FG02-96ER40963 (University of Tennessee). Experiments at NIST
were supported by the NSF through DMR-0086210 and DMR-9986442.


\end{document}